\documentclass[12pt]{article}
\pdfoutput =1
\usepackage{float} 
\usepackage{amsmath}
\usepackage{amsfonts}   
\usepackage{amssymb}

\begin{document}
\date{}
\title{{\bf{\Large Entanglement rules for holographic Fermi surfaces}}}
\author{
 {\bf {\normalsize Dibakar Roychowdhury}$
$\thanks{E-mail:  dibakarphys@gmail.com, dibakarr@iitk.ac.in}}\\
 {\normalsize  Indian Institute of Technology, Department of Physics,}\\
  {\normalsize Kanpur 208016, Uttar Pradesh, India}
}

\maketitle
\begin{abstract}
In this paper, based on the notion of Gauge/Gravity duality, we explore the laws of entanglement thermodynamics for most generic classes of Quantum Field Theories with hyperscaling violation. In our analysis, we note that for Quantum Field Theories with compressible \textit{quark} like excitation, the first law of entanglement thermodynamics gets modified due to the presence of an additional term that could be identified as the entanglement chemical potential associated with \textit{hidden} Fermi surfaces of the boundary theory. Most notably, we find that the so called entanglement chemical potential does not depend on the size of the entangling region and is purely determined by the quark d.o.f. encoded within the entangling region.
 \end{abstract}

\section{Overview and Motivation}
In the recent years, the holographic models for non relativistic Quantum Field Theories with hyperscaling violation \cite{Dong:2012se}-\cite{Kim:2012nb} has been found to provide an excellent theoretical framework in order to describe compressible states of quark matter (distributed over the so called \textit{hidden} Fermi surfaces) at strong coupling \cite{Ogawa:2011bz}-\cite{Shaghoulian:2011aa}. It turns out that for theories with holographic metals \cite{Liu:2009dm}-\cite{Lee:2008xf}, the total charge density (associated with the boundary theory) largely dominates over that of the volume enclosed by the Fermi surfaces. In \cite{Huijse:2011hp}-\cite{Sachdev:2010um}, the authors argue that such a deficit could be made up by considering hidden Fermi surfaces of fractionalized \textit{deconfined} charged fermionic excitation known as \textit{quarks}. 

In a recent paper \cite{Ogawa:2011bz}, the authors first pointed out that the presence of such compressible states of quark matter could be identified by computing the holographic entanglement entropy (HEE) \cite{Ryu:2006bv}-\cite{Ryu:2006ef} for the boundary theory, which instead of showing the usual area divergence, exhibits the so called \textit{logarithmic} divergence. These arguments were further sharpened by the authors in \cite{Huijse:2011ef}, who showed that an emerging infra-red geometry with arbitrary dynamic exponent ($ z $) and hyperscaling violating parameter ($ \theta $) precisely characterizes the presence of such compressible quark matter excitation distributed over the hidden Fermi surfaces of the boundary theory. 

The most natural question that arises in this context is whether there exists any notion of so called entanglement thermodynamics \cite{Bhattacharya:2012mi}-\cite{He:2013rsa} for small subsystems associated with these hidden Fermi surfaces of compressible quark excitation. In other words, whether one could possibly write down some version of the first law of entanglement thermodynamics in the presence of these fractionalized charged fermionic excitation. Now, in the presence of these additional quantum numbers (fermionic d.o.f.), the first law of entanglement thermodynamics must be modified in order to include a term which should be analogous to that of the chemical potential term associated with the standard first law of thermodynamic. The question therefore turns out to be whether one could define such an entity (which we call entanglement chemical potential) associated with hidden Fermi surfaces that leads to a modified first law of entanglement thermodynamics and if such an entity exits then whether it is universal in the same sense as that of the entanglement temperature. The purpose of the present article is to provide a systematic answer to these questions based on some concrete holographic computations.

The organisation of the paper is the following: In Section 2, we provide details regarding the gravitational set up in the bulk. In Section 3, we explore the modified first law of entanglement thermodynamics and compute the entanglement chemical potential associated with compressible states of quark matter. Finally, we conclude in Section 4.

\section{The background}
We start our analysis with a formal description to the gravitational solution in the bulk spacetime. The action that typically one considers is of the following form \cite{Dehghani:2015gza},
\begin{eqnarray}
S&=& S_{EH}+S_{M}\nonumber\\
S_{EH}&=&\frac{1}{16 \pi G_N}\int_{\mathcal{M}} d^{d+2}x\sqrt{-g} R + \frac{1}{8\pi G_N}\int_{\partial \mathcal{M}}d^{d+1}x \sqrt{-h}K\nonumber\\
S_{M}&=&\frac{1}{16 \pi G_N} \int_{\mathcal{M}} d^{d+2}x\sqrt{-g}\left[-\frac{1}{2}(\partial \Phi )^{2}+V(\Phi) -\frac{e^{\lambda_{1}\Phi}}{4}H^{2}+\frac{e^{\lambda_{2}\Phi}}{4}(-F^{2})^{s}\right] 
\label{e1}
\end{eqnarray}
where, $ H^{2}=H_{mn}H^{mn} $ and $ F^{2}=F_{mn}F^{mn} $ are the field strength tensors corresponding to two abelian one forms $ B_{m} $ and $ A_{m} $ respectively. The first abelian two form ($ H_{mn} $) coupled with the dilaton ($ \Phi $) generates the desired asymptotic solution of the Lifshitz type. On the other hand, the second Maxwell field ($ F_{mn} $) gives rise to the non linear charged Lifshitz black brane configuration. Here, $ V(\Phi) (= -2\Lambda e^{\gamma\Phi})$ is the so called exponential potential for the dilaton. 

To proceed further, in our analysis we set $ d=2 $. With this the resulting  expression for (non linear) charged hyperscaling violating Lifshitz configuration turns out to be\footnote{Here, $ \theta $ is the hyperscaling violating exponent and $ z $ is the so called dynamic critical exponent. For the moment we skip details regarding the gauge fields as well as the dilation profile as they are not directly required for the present analysis. Interested reader may have a look for their details in \cite{Dehghani:2015gza}.}\cite{Dehghani:2015gza},
\begin{eqnarray}
ds^{2}&=&r^{-\theta}\left( -r^{2z}f(r)dt^{2}+\frac{dr^{2}}{r^{2}f(r)}+r^{2}(dx^{2}+dy^{2})\right)\nonumber\\
f(r)&=&1-\frac{\mathcal{M}}{r^{z+2-\theta}}+\frac{\mathfrak{K}\mathcal{Q}^{\frac{2s}{2s-1}}}{r^{\zeta + 2 +z -\theta}},~~\zeta = z-2+\frac{2-\theta}{2s-1}\nonumber\\
\mathfrak{K}&=&\frac{(2s-1)r_{0}^{2(z-1-\theta / 2)}}{4(2-\theta)\zeta}2^{\frac{s}{2s-1}}(16 \pi)^{\frac{2s}{2s-1}}
\label{e2}
\end{eqnarray}
where, $ \mathcal{M} $ is the mass and $ \mathcal{Q} $ is the $ U(1) $ charge of the black brane.

The Hawking temperature \cite{Dehghani:2015gza} corresponding to the above black brane configuration (\ref{e2}) turns out to be,
\begin{eqnarray}
T_{H}=\frac{(2+z-\theta)r_H^{z}}{4 \pi}\left[ 1-\frac{\zeta \mathfrak{K}\mathcal{Q}^{2s/2s-1}}{2+z-\theta}r_H^{-(2+\zeta +z -\theta)}\right].
\label{e3}
\end{eqnarray}

On the other hand, the entropy density of the black brane turns out to be,
\begin{eqnarray}
s= \frac{r_H^{2-\theta}}{4 G_N}.
\label{e4}
\end{eqnarray}

Combining (\ref{e3}) and (\ref{e4}), the energy density \cite{Dehghani:2015gza} finally turns out to be,
\begin{eqnarray}
\varepsilon = \frac{(2-\theta)r_H^{2+z-\theta}}{16 \pi G_N}\left( 1+\frac{ \mathfrak{K}\mathcal{N}^{2s/2s-1}}{\mathcal{V}_{2}^{2s/2s-1}}r_H^{-(2+\zeta +z -\theta)} \right)
\label{E5} 
\end{eqnarray}
where, we have used the fact that the charge (density) of the black brane is dual to the \textit{quark} number density ($ \mathcal{N}/\mathcal{V}_{2} $) for the boundary field theory \cite{Dehghani:2015gza}.

Before we proceed further, let us first adopt some suitable radial coordinate (variable) namely, $ u=r_H/r $ such that the horizon of the black brane is placed at $ u=1 $ and the boundary of the space time is located near $ u \rightarrow 0 $.
With this choice of coordinates, the corresponding black brane metric (\ref{e2}) turns out to be,
\begin{eqnarray}
ds^{2}&=&\left( \frac{u}{r_H}\right)^{\theta} \left(-\left( \frac{r_H}{u}\right)^{2z}f(u)dt^{2}+\frac{du^{2}}{u^{2}f(u)}+\frac{r_H^{2}}{u^{2}}(dx^{2}+dy^{2})  \right)\nonumber\\
f(u)&=& 1- \mathcal{M}\left( \frac{u}{r_{H}}\right)^{z+2-\theta}+\mathfrak{K}\mathcal{Q}^{\frac{2s}{2s-1}}\left( \frac{u}{r_{H}}\right) ^{\zeta + 2 +z -\theta}
\label{e6}
\end{eqnarray}
which should be regarded as the starting point for our subsequent analysis \footnote{Although throughout this paper we would try to keep our analysis to be most generic, however, since the ultimate goal of our analysis is to explore the entanglement thermodynamics associated with the (hidden) Fermi surfaces of the boundary theory, therefore, in our analysis we would finally set, $ \theta =d-1=1 $ \cite{Ogawa:2011bz}-\cite{Huijse:2011ef} along with, $ z=2 $ and $ s=1 $. Our choice of $ s $ precisely corresponds to the fact that we are dealing with the linear Maxwell theory instead that of the Power Maxwell theory which we had started with. However, one could in principle choose other values of $ s $ as well and that should not affect any of the physical consequences of the theory.}. 
\section{Entanglement thermodynamics}
The holographic definition of entanglement entropy could be formally expressed as \cite{Ryu:2006bv},
\begin{eqnarray}
\mathcal{S}_{A}=\frac{Area(\gamma_{A})}{4 G_N}
\end{eqnarray}
where, $ \gamma_{A} $ is the measure of the minimal area of the surface that ends on the boundary.

In order to compute the holgraphic entanglement entropy (HEE), we consider our subsystem to be a strip of size, $ \mathcal{V}_{2}=L l $ where, $ 0<x< l$ and $ -L/2<y<L/2 $ such that, $ l \ll L $. We parametrize the minimal surface by, $ x=x(u) $ which yields the area of the extremal surface,
\begin{eqnarray}
Area=\frac{2L}{r_H} \int_{\epsilon}^{u_T}du ~\left( \frac{r_H}{u}\right)^{2-\theta}\sqrt{\frac{1}{f(u)(1-(u/u_T)^{2(2-\theta)})}}
\label{e9}
\end{eqnarray}
as well as the length of the strip as,
\begin{eqnarray}
l=\frac{2}{r_H} \int_{0}^{u_T}du ~\left( \frac{u}{u_T}\right)^{2-\theta} ~\sqrt{\frac{1}{f(u)(1-(u/u_T)^{2(2-\theta)})}}
\label{e10}
\end{eqnarray}
where, $ u=u_T $ corresponds to the turning point of the extremal surface ($ \gamma_{A} $).

\subsection{Case I: $ \mathcal{Q}=0 $}
In order to proceed further we first consider the uncharged case namely, $ \mathcal{Q}=0 $. Under such circumstances, turning on $ \mathcal{M} $ essentially corresponds to thermal excitation about the Fermi sea. Considering $ l $ to be very small (i.e, $ u_T/r_H \ll 1 $) and expanding $ f(u) $ perturbatively in $ \mathcal{M} $, from (\ref{e10}) we find,
\begin{eqnarray}
\frac{ l}{2}=\frac{\sqrt{\pi}u_T \Gamma\left(\frac{3-\theta}{2(2-\theta)} \right) }{ 2r_H(2-\theta)\Gamma\left(\frac{5-2 \theta}{2(2-\theta)} \right)}+\frac{\sqrt{\pi}  \mathcal{M}  u_T^{z+3-\theta}}{4(2-\theta) r_H^{z+3-\theta}}\frac{\Gamma\left(\frac{5+z-2 \theta}{2(2-\theta)} \right)}{\Gamma\left(\frac{7+z-3 \theta}{2(2-\theta)} \right)}.
\label{E10}
\end{eqnarray}

On the other hand, from (\ref{e9}) we find,
\begin{eqnarray}
\mathcal{S}_{A}=\mathcal{S}_{A}^{(0)}+\Delta\mathcal{S}_{A}
\end{eqnarray}
where, $ \mathcal{S}_{A}^{(0)}\sim \log (\epsilon/l) $ is the usual Log divergence associated with the hidden Fermi surfaces where $ \epsilon $ is the so called UV cut-off of the theory \cite{Ogawa:2011bz}. On the other hand, the change in HEE turns out to be,
\begin{eqnarray}
\Delta\mathcal{S}_{A}=\frac{\sqrt{\pi}\mathcal{M}L u_T^{z+1}}{8 G_N(2-\theta) r_H^{z+1}}\frac{\Gamma\left(\frac{1+z}{2(2-\theta)} \right)}{\Gamma\left(\frac{3+z- \theta}{2(2-\theta)} \right)}
\end{eqnarray}
which by virtue of (\ref{E10}) could be formally expressed as,
\begin{eqnarray}
\Delta\mathcal{S}_{A} \approx \frac{\mathcal{M}Ll^{z+1}(2-\theta)^{z}}{8 G_N (\sqrt{\pi})^{z}}\frac{\Gamma\left(\frac{1+z}{2(2-\theta)} \right)}{\Gamma\left(\frac{3+z- \theta}{2(2-\theta)} \right)}\left[ \frac{\Gamma\left(\frac{5-2 \theta}{2(2-\theta)} \right)}{\Gamma\left(\frac{3-\theta}{2(2-\theta)} \right)}\right]^{z+1}.
\label{E13} 
\end{eqnarray}

Using (\ref{E5}) and (\ref{E13}), the entanglement temperature finally turns out to be,
\begin{eqnarray}
T_{E}=\frac{\Delta \mathcal{E}}{\Delta \mathcal{S}_{A}}=\frac{\mathcal{V}_{2}\Delta\varepsilon}{\Delta \mathcal{S}_{A}}=\mathfrak{c}l^{-z}
\label{E14}
\end{eqnarray}
where,
\begin{eqnarray}
\mathfrak{c}=\frac{\pi^{z/2 -1}}{2(2-\theta)^{z-1}}\left[ \frac{\Gamma\left(\frac{1+z}{2(2-\theta)} \right)}{\Gamma\left(\frac{3+z- \theta}{2(2-\theta)} \right)}\right]^{-1} \left[ \frac{\Gamma\left(\frac{5-2 \theta}{2(2-\theta)} \right)}{\Gamma\left(\frac{3-\theta}{2(2-\theta)} \right)}\right]^{-(z+1)}.
\end{eqnarray}

Finally, for $ \theta =d-1=1 $ and, $ z=2 $ the entanglement temperature associated with the so called hidden Fermi surfaces could be trivially found to be,
\begin{eqnarray}
T_{E}^{(F)}=\mathfrak{c}^{(F)}l^{-2}
\label{E12}
\end{eqnarray}
where,
\begin{eqnarray}
\mathfrak{c}^{(F)}=\frac{4}{(\sqrt{\pi})^{3}}.
\end{eqnarray}

Eq.(\ref{E12}) precisely determines the \textit{universal} entanglement temperature associated with so called Fermi surfaces of the boundary theory at strong coupling and zero chemical potential. The value of the universal constant $ \mathfrak{c}^{(F)} $ is fixed by the shape of the entangling region, which for the present example turns out to be the thin rectangular strip. 

In the next subsection, we move on to the scenario with finite chemical potential which allows us to consider situations like exciting the system by creating particles even at zero temperature and therefore to check the validity of the laws of entanglement thermodynamics associated with Fermi surfaces at finite chemical potential.

\subsection{Case II: $ \mathcal{Q}\neq 0 $}
In this section, we explore the entanglement thermodynamics associated with (hidden) Fermi surfaces in the presence of additional conserved (global $ U(1) $) charges and/or finite number density of particles in the theory. There could be two possible scenarios under such circumstances namely, one could consider the change in entanglement entropy due to the change in temperature at finite charge density and vice versa. In the following we consider them one by one. 

We first compute the change in the HEE due to the change in temperature ($ T_H $) at fixed quark number density ($ \mathcal{N}_{A} $) encoded within the subsystem ($ A $). To do that, we first write down the entity, $ \Delta\mathcal{S}_{A} $ at finite temperature and charge density,
\begin{eqnarray}
\Delta\mathcal{S}_{A}(T_H,\mathcal{N}_{A}) \approx \frac{L r_H^{z+2-\theta}l^{z+1}(2-\theta)^{z}}{8 G_N(\sqrt{\pi})^{z}}\frac{\Gamma\left(\frac{1+z}{2(2-\theta)} \right)}{\Gamma\left(\frac{3+z- \theta}{2(2-\theta)} \right)}\left[ \frac{\Gamma\left(\frac{5-2 \theta}{2(2-\theta)} \right)}{\Gamma\left(\frac{3-\theta}{2(2-\theta)} \right)}\right]^{z+1}\left( 1+\frac{ \mathfrak{K}\mathcal{N}_{A}^{2}}{\mathcal{V}_{2}^{2}}r_H^{-2(1+z -\theta)} \right)\nonumber\\
-\frac{L l^{2z+1-\theta}(2-\theta)^{2z-\theta}\mathfrak{K}\mathcal{N}_{A}^{2}}{8 G_N\mathcal{V}_{2}^{2}(\sqrt{\pi})^{2z-\theta}}\frac{\Gamma\left(\frac{1+2z-\theta}{2(2-\theta)} \right)}{\Gamma\left(\frac{3+2z- 2\theta}{2(2-\theta)} \right)}\left[ \frac{\Gamma\left(\frac{5-2 \theta}{2(2-\theta)} \right)}{\Gamma\left(\frac{3-\theta}{2(2-\theta)} \right)}\right]^{2z+1-\theta}.
\label{E18} 
\end{eqnarray}

Next, we note down the corresponding change at zero temperature and finite particle density namely,
\begin{eqnarray}
\Delta\mathcal{S}_{A}(T_H=0,\mathcal{N}_{A}) \approx \frac{L r_H^{(0)z+2-\theta}l^{z+1}(2-\theta)^{z}}{8 G_N(\sqrt{\pi})^{z}}\frac{\Gamma\left(\frac{1+z}{2(2-\theta)} \right)}{\Gamma\left(\frac{3+z- \theta}{2(2-\theta)} \right)}\left[ \frac{\Gamma\left(\frac{5-2 \theta}{2(2-\theta)} \right)}{\Gamma\left(\frac{3-\theta}{2(2-\theta)} \right)}\right]^{z+1}\left( 1+\frac{ \mathfrak{K}\mathcal{N}_{A}^{2}}{\mathcal{V}_{2}^{2}}r_H^{-(0)2(1+z -\theta)} \right)\nonumber\\
-\frac{L l^{2z+1-\theta}(2-\theta)^{2z-\theta}\mathfrak{K}\mathcal{N}_{A}^{2}}{8G_N\mathcal{V}_{2}^{2}(\sqrt{\pi})^{2z-\theta}}\frac{\Gamma\left(\frac{1+2z-\theta}{2(2-\theta)} \right)}{\Gamma\left(\frac{3+2z- 2\theta}{2(2-\theta)} \right)}\left[ \frac{\Gamma\left(\frac{5-2 \theta}{2(2-\theta)} \right)}{\Gamma\left(\frac{3-\theta}{2(2-\theta)} \right)}\right]^{2z+1-\theta}.
\label{E19} 
\end{eqnarray}

Using (\ref{E18}) and (\ref{E19}), we finally obtain the change in entanglement variation (at fixed particle number density) as,
\begin{eqnarray}
\Delta\mathcal{S}_{A}|_{\mathcal{N}_{A}}&=& \Delta\mathcal{S}_{A}(T_H,\mathcal{N}_{A})-\Delta\mathcal{S}_{A}(T_H=0,\mathcal{N}_{A})\nonumber\\
&=&\frac{L l^{z+1}(2-\theta)^{z}}{8G_N(\sqrt{\pi})^{z}}\frac{\Gamma\left(\frac{1+z}{2(2-\theta)} \right)}{\Gamma\left(\frac{3+z- \theta}{2(2-\theta)} \right)}\left[ \frac{\Gamma\left(\frac{5-2 \theta}{2(2-\theta)} \right)}{\Gamma\left(\frac{3-\theta}{2(2-\theta)} \right)}\right]^{z+1}\Xi (r_H)
\label{E20}
\end{eqnarray}
where,
\begin{eqnarray}
\Xi (r_H)=r_H^{z+2-\theta}\left( 1+\frac{ \mathfrak{K}\mathcal{N}_{A}^{2}}{\mathcal{V}_{2}^{2}}r_H^{-2(1+z -\theta)} \right)-r_H^{(0)z+2-\theta}\left( 1+\frac{ \mathfrak{K}\mathcal{N}_{A}^{2}}{\mathcal{V}_{2}^{2}}r_H^{-(0)2(1+z -\theta)} \right).
\end{eqnarray}

Similar calculations for the energy density (\ref{E5}) yields,
\begin{eqnarray}
\Delta \varepsilon |_{\mathcal{N}_{A}} = \frac{(2-\theta)}{16 \pi G_N}\Xi (r_H).
\label{E22}
\end{eqnarray}

Eqs.(\ref{E20}) and (\ref{E22}) could be combined together in order to yield the first law of Entanglement thermodynamics namely,
\begin{eqnarray}
\Delta \mathcal{E} |_{\mathcal{N}_{A}}=T_E ~~ \Delta\mathcal{S}_{A} |_{\mathcal{N}_{A}}
\end{eqnarray}
where $ T_E $ is the entanglement temperature as defined above in (\ref{E14}).

We next move on to the second scenario where one could possibly have a variation in the entanglement entropy due to the variation in the particle number within the system itself. To do that, as first step of our analysis, from (\ref{e3}) we note that,
\begin{eqnarray}
\Delta T_H =0\Rightarrow \Delta r_H =\frac{2(z-\theta)\mathfrak{K}\mathcal{N}_{A}r_H^{-(1+2z-2 \theta)}}{z\mathcal{V}_{2}^{2}(2+z-\theta)\left(1+\frac{(z-\theta)\mathfrak{K}\mathcal{N}_{A}^{2}(2+z-2\theta)}{z \mathcal{V}_{2}^{2}(2+z-\theta)} r_H^{-2(1-\theta +z)}\right) }\Delta \mathcal{N}_{A}.
\label{E24}
\end{eqnarray}

Using (\ref{E24}), from (\ref{E18}) we finally obtain,
\begin{eqnarray}
\frac{\Delta\mathcal{S}_{A}|_{T_H}}{\Delta \mathcal{N}_{A}}=\left[ \frac{\Gamma\left(\frac{5-2 \theta}{2(2-\theta)} \right)}{\Gamma\left(\frac{3-\theta}{2(2-\theta)} \right)}\right]^{z+1}\left[ \frac{\Theta\Gamma\left(\frac{1+z}{2(2-\theta)} \right)}{\Gamma\left(\frac{3+z- \theta}{2(2-\theta)} \right)} -\frac{ l^{2z-1-\theta}(2-\theta)^{2z-\theta}\mathfrak{K}\mathcal{N}_{A}}{4 G_N L(\sqrt{\pi})^{2z-\theta}}\frac{\Gamma\left(\frac{1+2z-\theta}{2(2-\theta)} \right)}{\Gamma\left(\frac{3+2z- 2\theta}{2(2-\theta)} \right)}\left[ \frac{\Gamma\left(\frac{5-2 \theta}{2(2-\theta)} \right)}{\Gamma\left(\frac{3-\theta}{2(2-\theta)} \right)}\right]^{z-\theta} \right]\nonumber\\
\label{E25} 
\end{eqnarray}
where,
\begin{eqnarray}
\Theta =\frac{\mathfrak{K}\mathcal{N}_{A}(z-\theta) r_H^{-z+\theta}l^{z-1}(2-\theta)^{z}\left(1+\frac{z(\theta -z)\mathfrak{K}\mathcal{N}^{2}_{A}r_H^{-2(1-\theta +z)}}{L^{2}l^{2}(z+2-\theta)} \right) }{4 G_N L(\sqrt{\pi})^{z}z\left(1+\frac{(z-\theta)\mathfrak{K}\mathcal{N}_{A}^{2}(2+z-2\theta)}{z L^{2}l^{2}(2+z-\theta)} r_H^{-2(1-\theta +z)}\right)}
+\frac{l^{z-1}\mathfrak{K}\mathcal{N}_{A}(2-\theta)^{z}r_H^{-z+\theta}}{4G_N L (\sqrt{\pi})^{z}}.\nonumber\\
\end{eqnarray}

On the other hand, performing similar variations in (\ref{E5}) we find,
\begin{eqnarray}
\frac{\mathcal{V}_{2}\Delta\varepsilon |_{T_H}}{\Delta \mathcal{N}_{A}}=\frac{(2-\theta)}{16 \pi G_N}\left( \frac{2(z-\theta)\mathfrak{K}\mathcal{N}_{A}r_H^{-z+ \theta}\left(1+\frac{z(\theta -z)\mathfrak{K}\mathcal{N}^{2}_{A}r_H^{-2(1-\theta +z)}}{L^{2}l^{2}(z+2-\theta)} \right) }{zL l\left(1+\frac{(z-\theta)\mathfrak{K}\mathcal{N}_{A}^{2}(2+z-2\theta)}{z L^{2}l^{2}(2+z-\theta)} r_H^{-2(1-\theta +z)}\right) }+\frac{2\mathfrak{K}\mathcal{N}_{A}r_H^{-z+\theta}}{Ll}\right). 
\label{E27}
\end{eqnarray}

Using (\ref{E14}), one could in fact combine (\ref{E25}) and (\ref{E27}) into a nice identity, namely the first law of entanglement thermodynamics,
\begin{eqnarray}
T_E \Delta\mathcal{S}_{A}|_{T_H} = \Delta \mathcal{E}|_{T_H}-\mu_E \Delta \mathcal{N}_{A}
\end{eqnarray}
where, 
\begin{eqnarray}
\mu_E =\frac{\mathfrak{K}\mathcal{N}_{A}(2-\theta)^{z-\theta +1}}{8 \pi^{1+(z-\theta)/2}G_N L l^{1+\theta -z}}\frac{\Gamma\left(\frac{1+2z-\theta}{2(2-\theta)} \right)}{\Gamma\left(\frac{3+2z- 2\theta}{2(2-\theta)} \right)}\left[ \frac{\Gamma\left(\frac{5-2 \theta}{2(2-\theta)} \right)}{\Gamma\left(\frac{3-\theta}{2(2-\theta)} \right)}\right]^{z-\theta} \left[ \frac{\Gamma\left(\frac{1+z}{2(2-\theta)} \right)}{\Gamma\left(\frac{3+z- \theta}{2(2-\theta)} \right)}\right]^{-1}
\label{E29}
\end{eqnarray}
is the most general form of the entanglement chemical potential associated with holographic Lifshitz theories in the presence of hyperscaling violation. Eq.(\ref{E29}), therefore suggests that for generic hyperscaling violating theories (with arbitrary $ z $ and $ \theta $) the entanglement chemical potential is universal, namely it is determined by the size of the entangling region,
\begin{eqnarray}
\mu_E  \sim \mathcal{N}_{A} l^{z-\theta -1}.
\end{eqnarray}
However, we encounter a surprise for a theory with holographic (hidden) Fermi surfaces (with, $ z=2 $ and $ \theta =1 $), where we find that unlike the case for the entanglement temperature, the entanglement chemical potential does not at all depend on the size of the entangling region namely,
\begin{eqnarray}
\mu_{E}^{(F)}=\frac{\mathfrak{K}\mathcal{N}_{A}}{6 \pi^{2}G_N L}
\end{eqnarray}
which therefore suggests that the entanglement chemical potential associated with hidden Fermi surfaces of compressible quark matter is directly proportional to the total number of quark excitation encoded within the entangling region. Therefore, this leads to the following conclusion that for sufficiently small entangling regions, the entanglement chemical potential associated with (holographic) Fermi surfaces is independent that of the size of the entangling region and is determined solely due to the fractionalized fermionic d.o.f. associated with compressible quark matter distributed over the hidden Fermi surfaces.

\section{Summary and final remarks}
We now summarize the key findings of our analysis. In the present paper, we consider holographic models that describe compressible quark like excitation distributed over hidden Fermi surfaces of the boundary theory and explore the laws of entanglement thermodynamics associated with small subsystems in such a configuration. The upshot of our analysis is the following: We note that in the presence of fermionic excitation, one could associate quantities like entanglement chemical potential with that of the hidden Fermi surfaces of the theory and indeed write down a modified version of the usual entanglement thermodynamics. Throughout this paper we keep our analysis to be most generic. However, as a surprise, for small entangling regions, we note that the so called entanglement chemical potential associated with compressible quark excitation does not depend on the size of the entangling region and is solely determined by the total number of quark excitation encoded within that subsystem. This turns out to be the main result of our analysis and it clearly categories a theory with hidden Fermi surfaces into a separate universality class from that of the other non conformal theories those have been studied earlier in the literature \cite{Park:2015afa}.

{\bf {Acknowledgements :}}
 The author would like to acknowledge the financial support from UGC (Project No UGC/PHY/2014236).

\end{document}